%% file: eprint.tex
\documentclass[12pt]{article}
\usepackage{graphicx}

\newcommand\pubnumber{LHCb-PROC-2012-030}
\newcommand\pubdate{\today}

\def\oxford{Department of Physics, University of Oxford, OX1 3RH,
  United Kingdom}

\def\Title#1{\begin{center} {\Large #1 } \end{center}}
\def\Author#1{\begin{center}{ \sc #1} \end{center}}
\def\Address#1{\begin{center}{ \it #1} \end{center}}

\newcommand\pubblock{\rightline{\begin{tabular}{l} \pubnumber\\
         \pubdate  \end{tabular}}}
\newenvironment{Abstract}{\begin{quotation}  }{\end{quotation}}
\newenvironment{Presented}{\begin{quotation} \begin{center} 
             PRESENTED AT\end{center}\bigskip 
      \begin{center}\begin{large}}{\end{large}\end{center} \end{quotation}}
\def\Acknowledgements{\bigskip  \bigskip \begin{center} \begin{large}
             \bf ACKNOWLEDGEMENTS \end{large}\end{center}}

\input econfmacros.tex

\begin{document}
\begin{titlepage}
\pubblock

\vfill
\Title{Model-independent searches for CP violation in multi-body charm decays
 }
\vfill
\Author{ Hamish Gordon, for the LHCb Collaboration}
\Address{\oxford}
\vfill
\begin{Abstract}
Model-independent techniques for CP violation searches in multi-body charm decays
are discussed. Examples of recent analyses from BaBar and LHCb are used to illustrate the
experimental challenges involved.

\end{Abstract}
\vfill
\begin{Presented}
The 5th International Workshop on Charm Physics\\
(Charm 2012)\\
4-17 May 2012, Honolulu, Hawai'i 96822
\end{Presented}
\vfill
\end{titlepage}
\def\thefootnote{\fnsymbol{footnote}}
\setcounter{footnote}{0}

\section{Introduction}

The recent evidence for CP violation (CPV) in  singly Cabibbo suppressed $D^{0}$ decays to two-body
final states from LHCb~\cite{charmcpv} and
CDF~\cite{Collaboration:2012qw} has heightened interest from
theoreticians in charm physics. In order to investigate
these promising hints further, multi-body decays with three or
four particles in the final state will be
needed. Direct CPV arises when decays with two possible routes to the same
final state with different relative weak
and strong phases interfere. The intermediate resonances
in multi-body final states have different strong phases, and thus if
there are also different weak phases, CPV is
guaranteed. Moreover, in multi-body $D$ decays it is possible in principle to
extract information about these strong and weak phases which remains hidden
in two-body systems. 

Furthermore, some of the easiest charm decays to reconstruct
experimentally have three or four charged particles in the final
state. For example, the most abundant flavour-tagged singly Cabibbo
suppressed final state available to most experiments is the three-body $D^{+} \to
K^{-}K^{+}\pi^{+}$ decay \cite{pdg}, and so, naively, this should be the most
promising charm decay mode for CPV searches.

However, multi-body charm decays present several key challenges. To
exploit the phase information to the full requires an amplitude model. Such
models carry with them systematic uncertainties that are difficult to
quantify and depend on choices made by the analysts. The properties of the
many possible scalar resonances required by the amplitude models are
often poorly known, and model uncertainties now approach or even exceed statistical
uncertainties in many amplitude analyses (e.g. \cite{Aubert:2008ao}). Therefore it is
worth investigating model-independent techniques which are not subject to these
uncertainties. With the large data samples now under study at the
B-factories and LHCb,
such techniques should also be faster and simpler to
implement.

There are now several model-independent techniques in the
experimentalists' armoury. I will discuss recent analyses which
illustrate some of these well.

\section{$D^{0} \to hh\pi^{0}$ at BaBar}

One of the first model-independent searches for CPV was
performed at BaBar on the $D^{*+}$-tagged decays $D^{0} \to \pi^{+}\pi^{-}\pi^{0}$ and
$D^{0} \to K^{+}K^{-}\pi^{0}$ ~\cite{Aubert:2008yd}. An
amplitude analysis, a phase-space-integrated search for CPV, and two
model-independent techniques were employed. In $82,468\pm321$ $\pi^{+}\pi^{-}\pi^{0}$ and $11,278\pm110$
$K^{+}K^{-}\pi^{0}$ decays, no evidence of CPV is found.

In the first search, the Dalitz plots of these decays are divided
into simple grids of bins. In each bin, the variable $S_{CP}$, defined
as
\begin{equation}
S_{CP}^{i} =\frac{N^i_{D^{+}}-N^i_{D^{-}}\alpha}{\sqrt{(\Delta
    N^i_{D^{+}})^{2}+(\Delta N^i_{D^{-}})^{2}\alpha^{2}} }, \qquad
\alpha  = \frac{\sum_{i} N^i_{D^{+}}}{\sum_{i} N^i_{D^{-}}}
\end{equation}
is calculated. Here, $N^i_{D^{+}}$ represents the number of $D^{+}$
decays observed in a given bin $i$, $\Delta
    N^i_{D^{+}}$ its uncertainty, and $\alpha$ is an overall
normalisation used to cancel any production asymmetry effect. This is similar to the ``Miranda'' method~\cite{Bediaga:2009tr}.

\begin{figure}[htb]
\centering
\includegraphics[width=7cm]{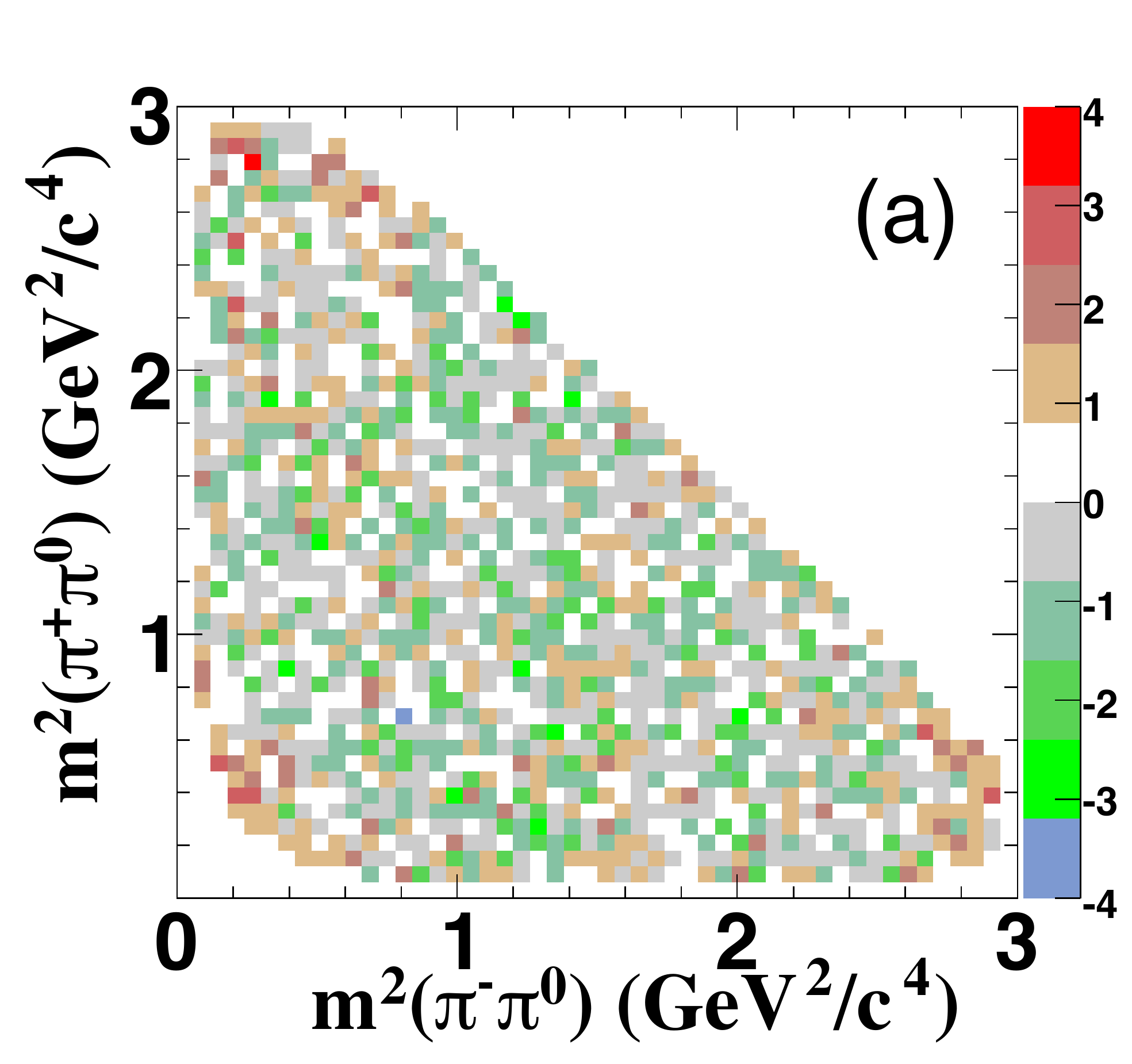}
\includegraphics[width=7cm]{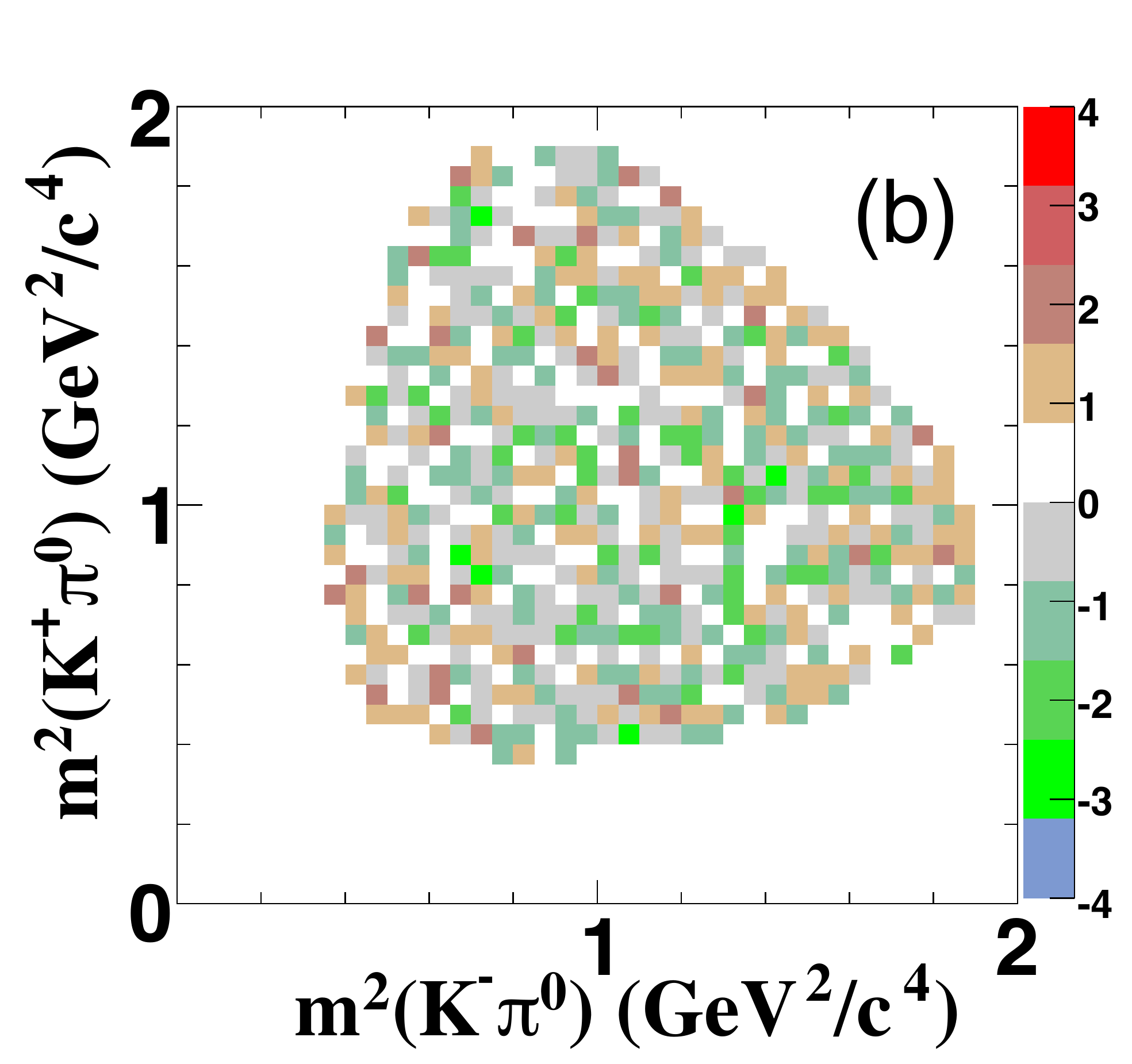}
\caption{Distribution of $S_{CP}^{i}$ values over the Dalitz plots:
  $\D^{0} \to \pi^{+}\pi^{-}\pi^{0}$ on the left, $\D^{0} \to
  K^{+}K^{-}\pi^{0}$ on the right.}
\label{fig:bb:scp}
\end{figure}

The sum of the squares of the $S_{CP}^{i}$ over all bins is a
$\chi^{2}$ for consistency with no CPV, with a number of
degrees of freedom equal to the number of bins minus one (due to the
overall normalisation). BaBar obtain a one-sided Gaussian
confidence level for consistency with no CPV of 32.8\% for
$\pi^{+}\pi^{-}\pi^{0}$ and 16.6\% for $K^{+}K^{-}\pi^{0}$. The
distribution of $S_{CP}^{i}$ values over the Dalitz plots is shown in
Figure~\ref{fig:bb:scp}.

BaBar also perform a complementary search for CPV which makes use of the
angular moments of the helicity angle $\theta_{H}$. This is defined for decays
of the type $D \to r(AB) C$ as the angle between the momentum
of $A$ in the $AB$ rest frame and the direction opposite
to the $D$ momentum in that same frame. The procedure is to bin the Dalitz plot projections
in two-body invariant mass, then weight the efficiency-corrected number
of events in a bin by the $l^{th}$ Legendre polynomial function of
$\cos\theta_{H}$. This yields moments 
\begin{equation}
 P_{0} = \frac{|S|^{2}+|P|^{2}}{\sqrt{2}}, \; P_{1} =
\sqrt{2}|S||P|\cos \theta_{SP}, \; P_{2} =
\sqrt{\frac{2}{5}}|P|^{2}, \; \ldots
\end{equation}
which allow one to separate the $S$ and $P$-wave components of the
Dalitz plot. An asymmetry variable $X_{l, CP}$ is then defined as 
\begin{equation}
X_{l, CP}=  \frac{\bar{P_{l}}-\alpha P_{l}}{\sqrt{\sigma^{2}_{\bar{P_{l}}} +
      \alpha^{2}\sigma_{P_{l}}^{2}}}
\end{equation}
with $\alpha$ defined as before. Legendre polynomial moments of orders
greater than eight are neglected. A $\chi^{2}$ for consistency with no
CPV is calculated in the same way, this time accounting for
correlations between the eight different angular moments $l$ with a
correlation coefficient $\rho_{ij}$ obtained from simulation:
\begin{equation}
\chi^{2} = \sum_{0}^{k} \sum_{i=0}^{7}\sum_{j=0}^{7}X_{i}\rho_{ij}X_{j}
\end{equation}
Here $k$ is the number of intervals of invariant mass and the number
of degrees of freedom is $8k$. BaBar find the one-sided confidence
level for no CPV to be 28.2\% for the $\pi^{+}\pi^{-}$, 28.4\% for the
$\pi^{+}\pi^{0}$, 63.1\% for the $K^+K^−$, and 23.8\% for the
$K^+\pi^0$ sub-systems. It should be noted that this procedure assumes
there is no interference from crossing channels and that only $S$ and
$P$ waves are present.

\section{$D^{+} \to K^{-}K^{+}\pi^{+}$ at LHCb}

\begin{figure}[htb]
\centering
\includegraphics[width=7cm]{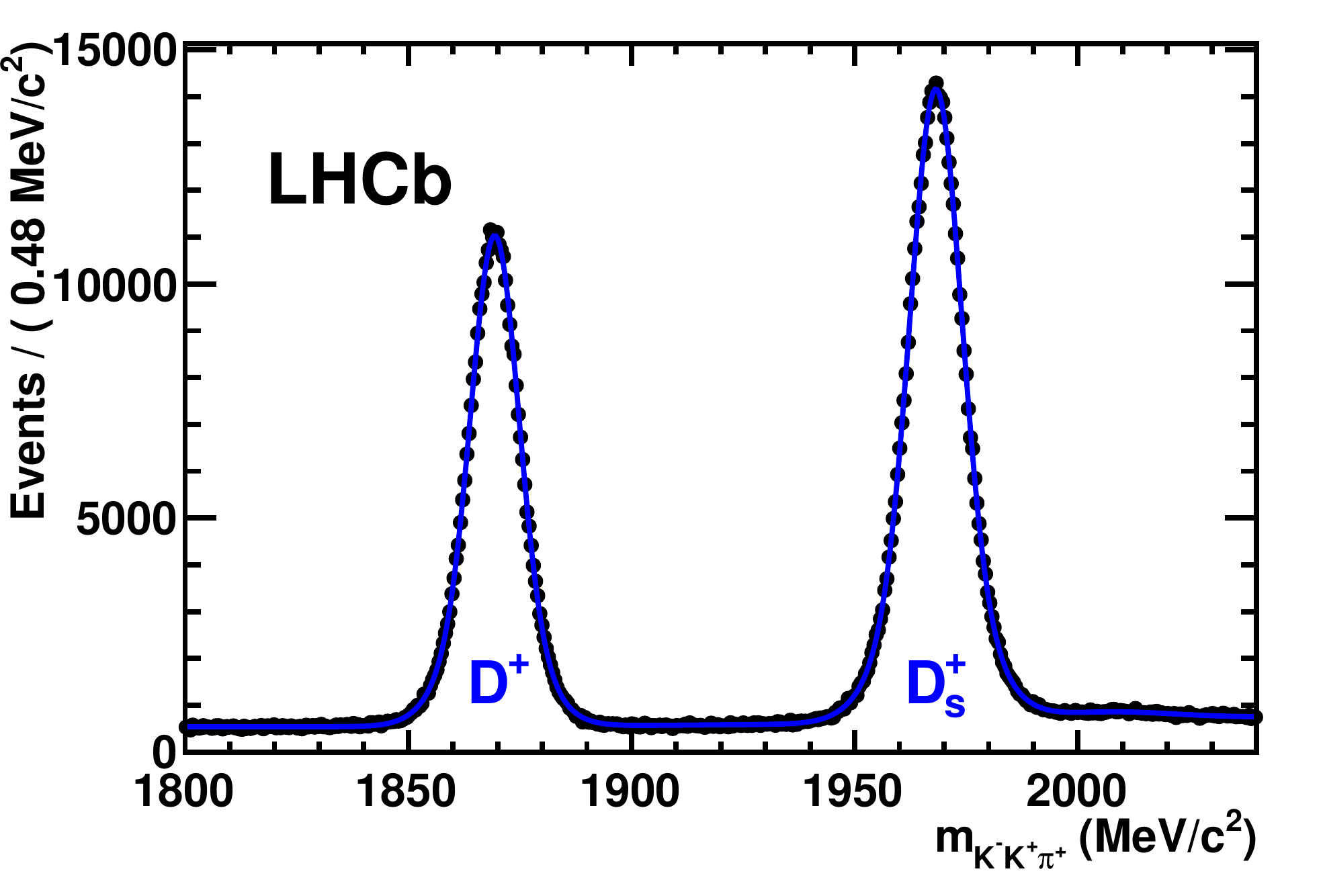}
\includegraphics[width=7cm]{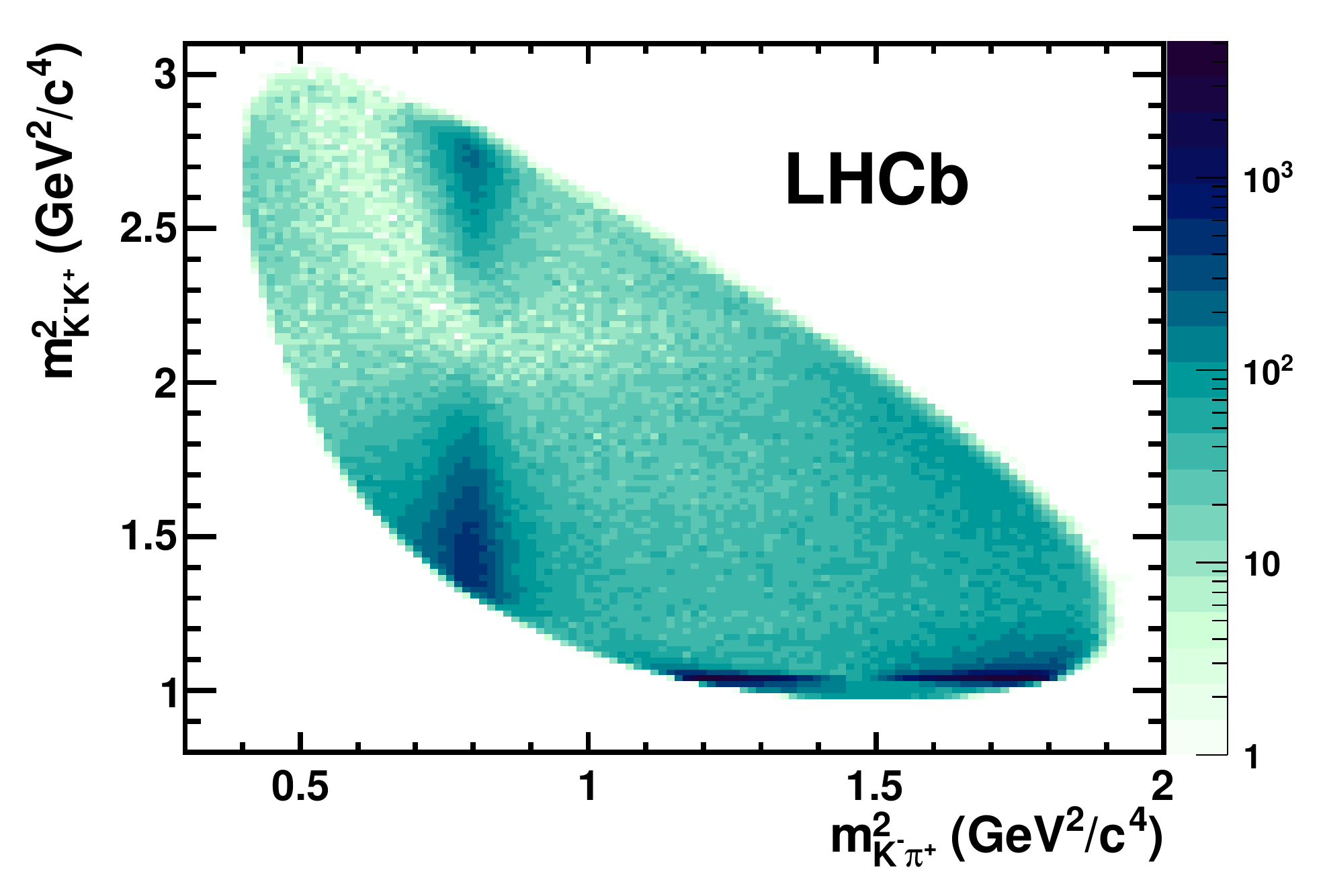}
\caption{$K^{-}K^+\pi^+$  Mass distribution (left) and  $D^{+} \to
  K^{-}K^+\pi^+$  Dalitz plot (right).}
\label{fig:lb:massdp}
\end{figure}

A similar analysis strategy was pursued with the first 35~pb$^{-1}$ of data
taken at the LHCb experiment in 2010~\cite{kkpiLHCb}. The hadronic environment and the higher statistics present some
additional challenges. For example, all multi-body analyses are
subject to some extent to the variation of interaction asymmetries of
particles with detector material across the Dalitz plot, because these
asymmetries can depend on momentum. At LHCb the high boost tends to
reduce these problems. This was demonstrated in the convenient Cabibbo-favoured
control channel, $D^{+}_{s} \to K^{-}K^+\pi^+$, which has similar
kinematics and signal yield to the $D^{+}$ decay.
The reconstructed $K^{-}K^+\pi^+$ mass
distribution and  $D^{+} \to K^{-}K^+\pi^+$ Dalitz plot is shown in Figure \ref{fig:lb:massdp}. There are around
370,000 $D^{+} \to K^{-}K^+\pi^+$ signal events with 90\% purity in
the 2010 dataset.

Monte Carlo simulations based on an amplitude model from CLEO-c~\cite{Rubin:2008aa} are used
to develop binning schemes for the Dalitz plot with greater sensitivity
to plausible types of CP violation than a simple regular grid. For
example, it is found that the strong phase can change across
resonances. If one bin is used for the whole resonance, the CPV would
change sign within it and the $S_{CP}$ value for that bin would
probably be consistent with 0. Therefore, without sacrificing model
independence, the resonances are divided up into bins an appropriate way to
account for the changing strong phase. With 25 bins in the Dalitz
plot, the Monte Carlo studies suggest, for example, that a CP-violating
phase in the $\phi(1020)$ of $5^{\circ}$ would be observed at the $3\sigma$
C.L. with $\sim 90\%$ probability. Other possible CPV signals are
explored and, for example, the method is similarly sensitive to an
$\sim 11\%$ CPV in the magnitude of the $\kappa(800)$.

The charge asymmetries in the $D^{+}_{s} \to K^{-}K^+\pi^+$ control
channel, and also in the Cabibbo-favoured $D^{+}\to K^{-}\pi^+\pi^+$ decay, are
investigated to eliminate the possibility of observing detector
asymmetries as signals of CPV. No significant charge asymmetries are
observed, indicating that no Dalitz-plot
dependent fake asymmetries are yet detectable.

\begin{figure}[htb]
\centering
\includegraphics[width=3.5cm]{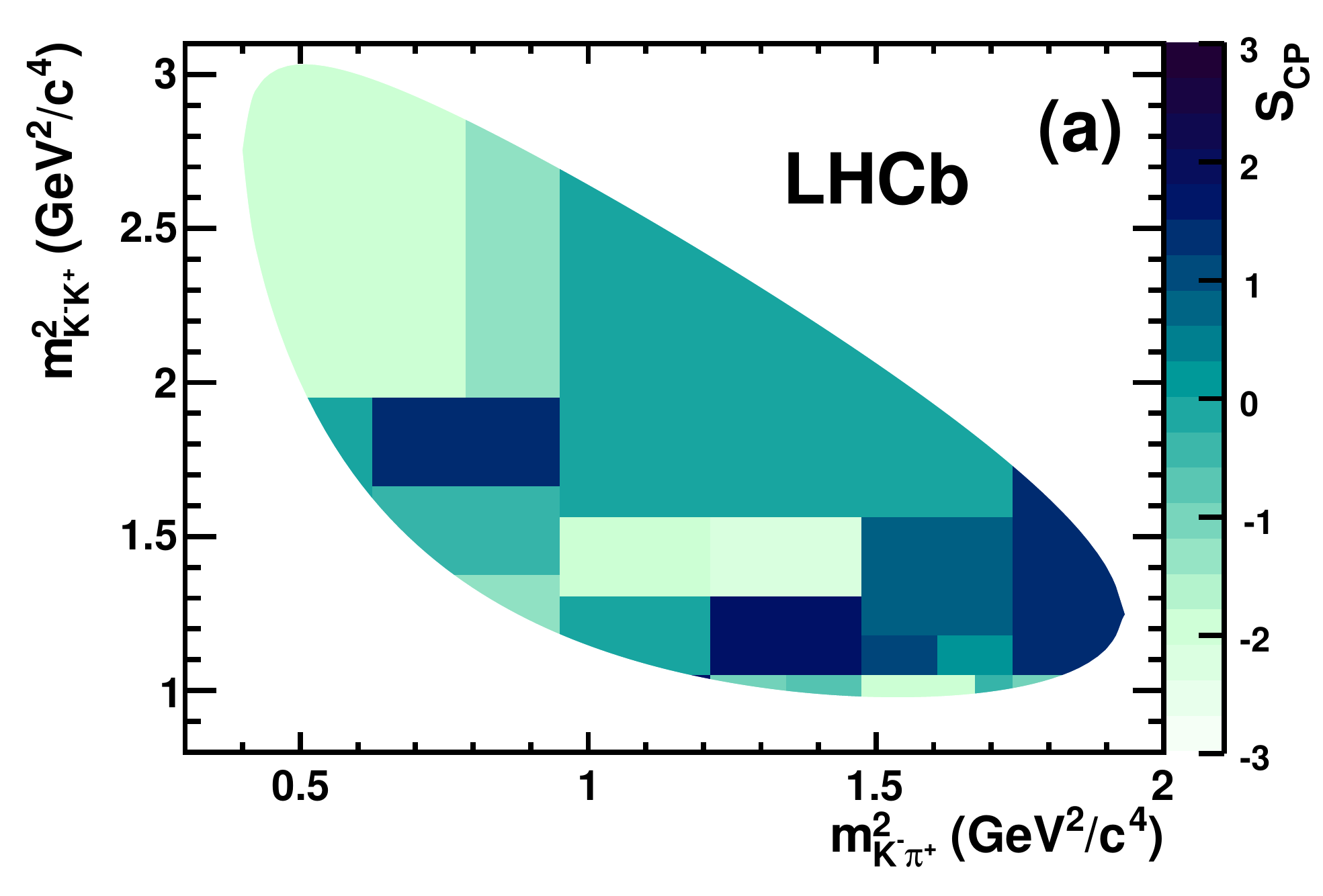}
\includegraphics[width=3.5cm]{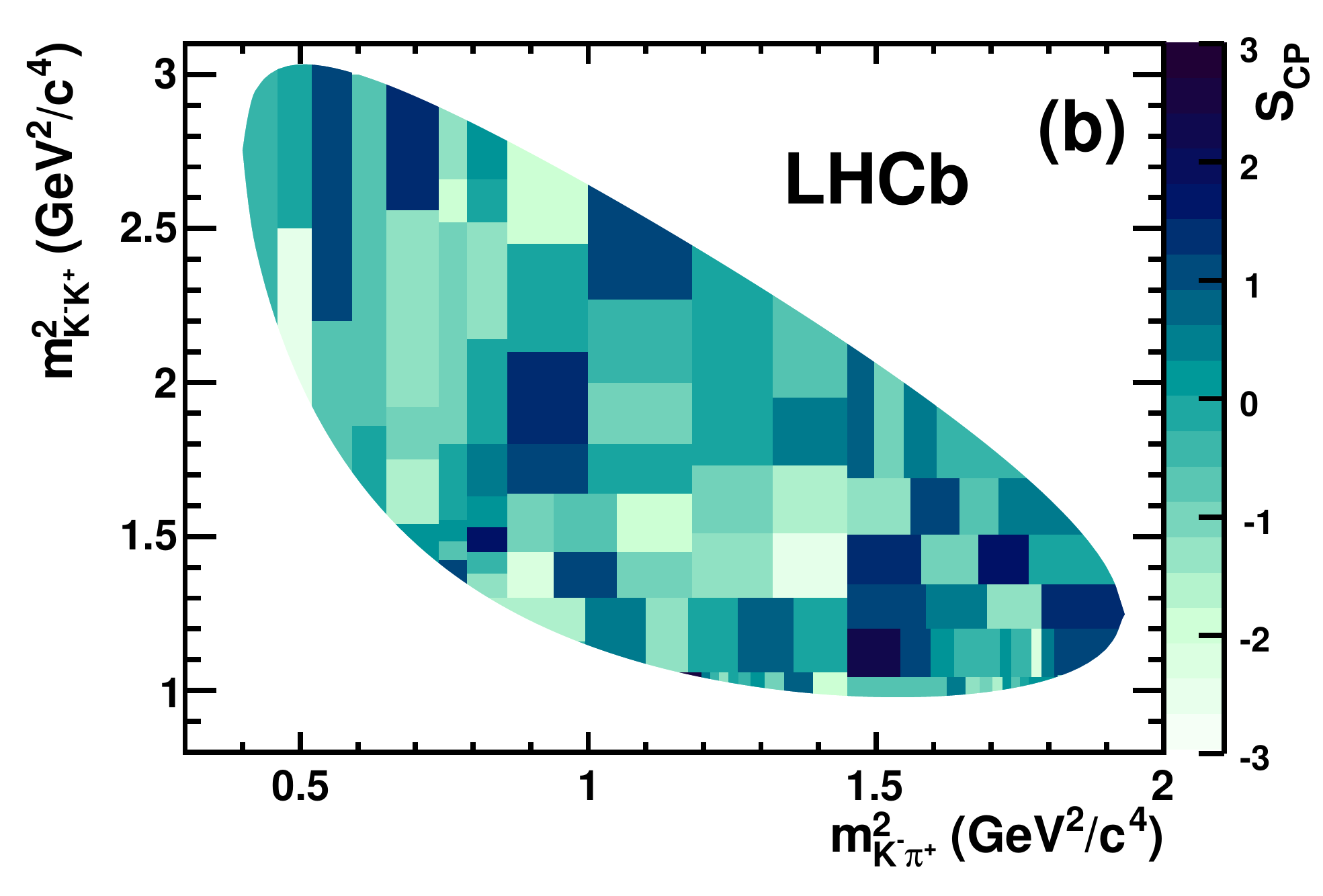}
\includegraphics[width=3.5cm]{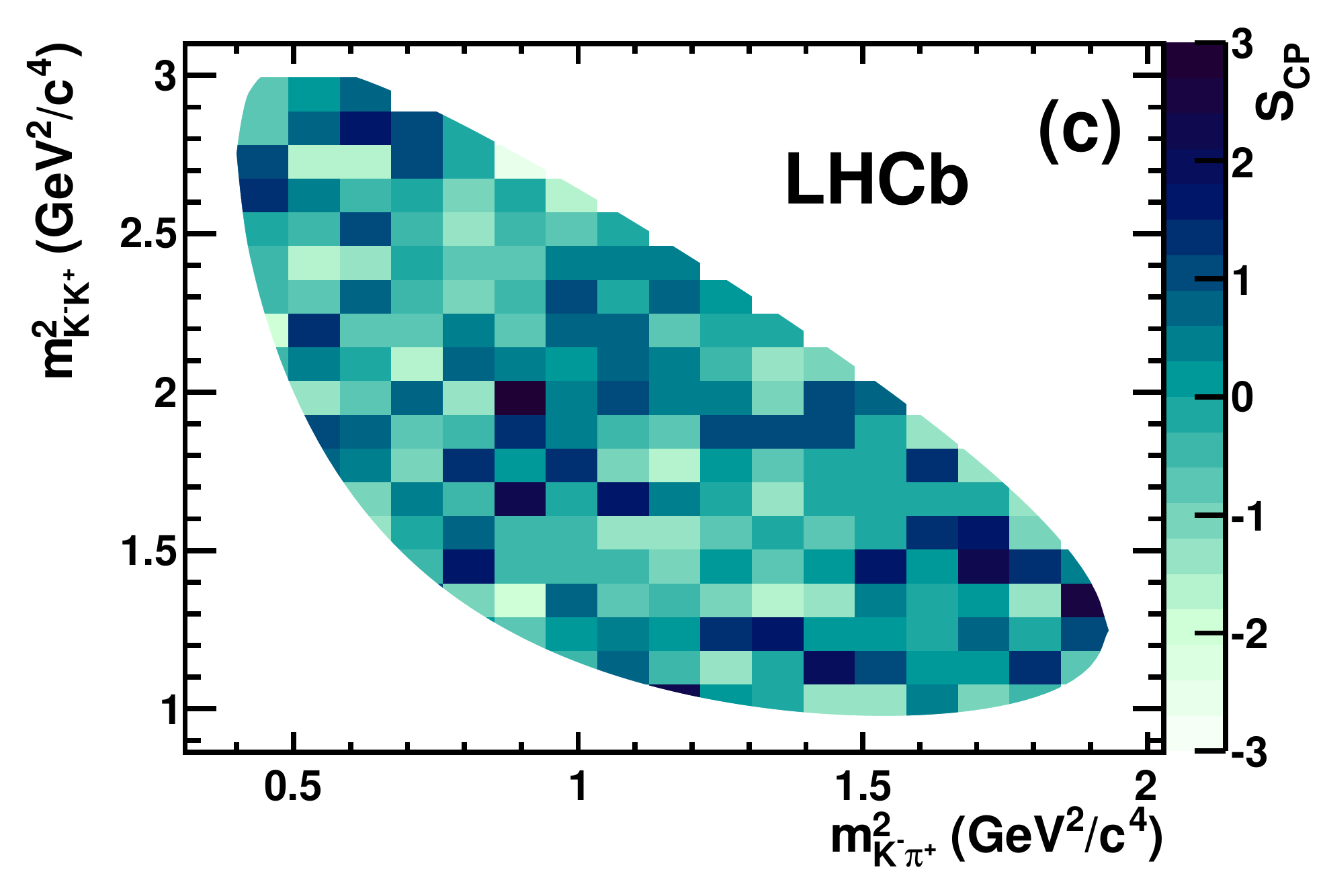}
\includegraphics[width=3.5cm]{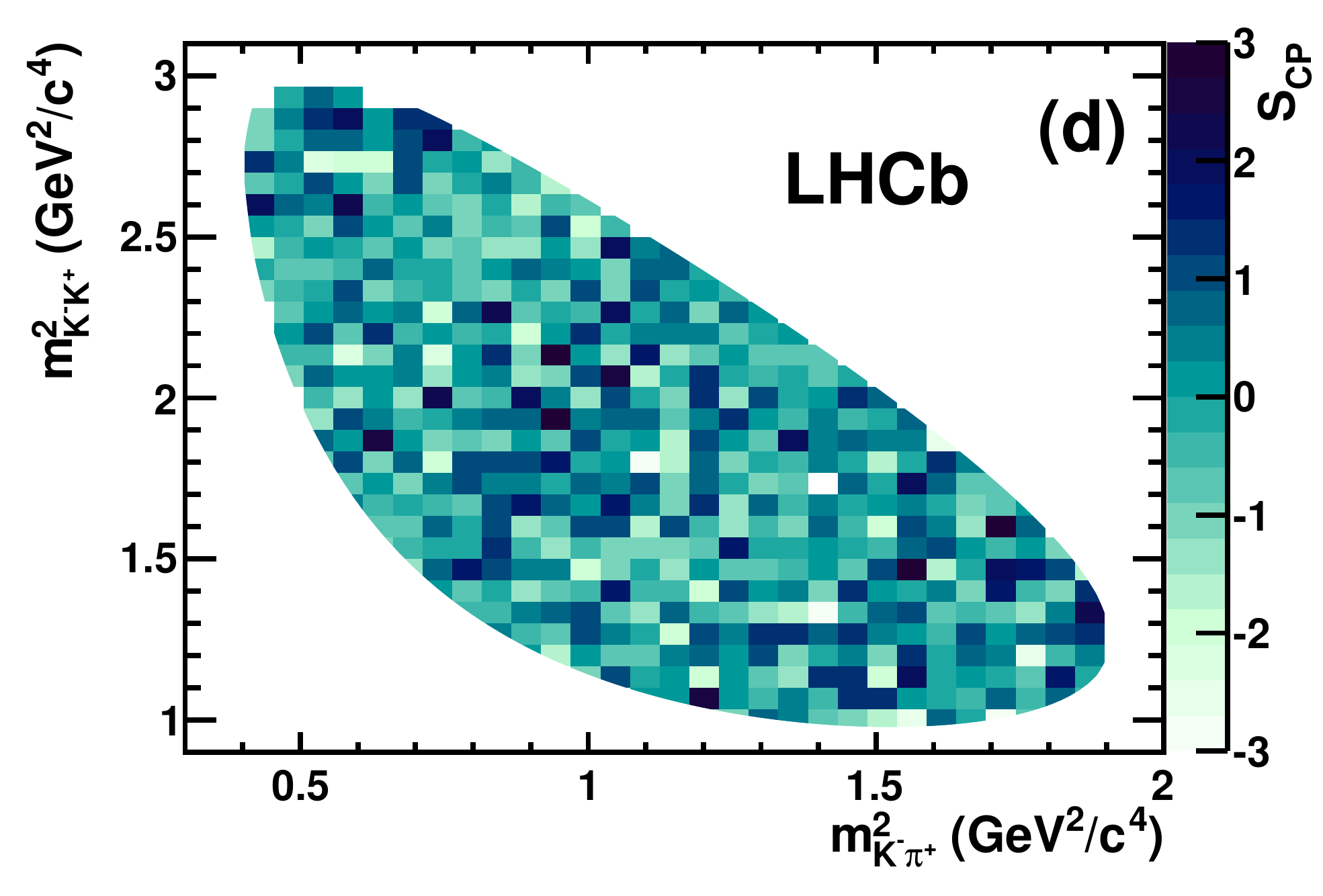}

\includegraphics[width=3.5cm]{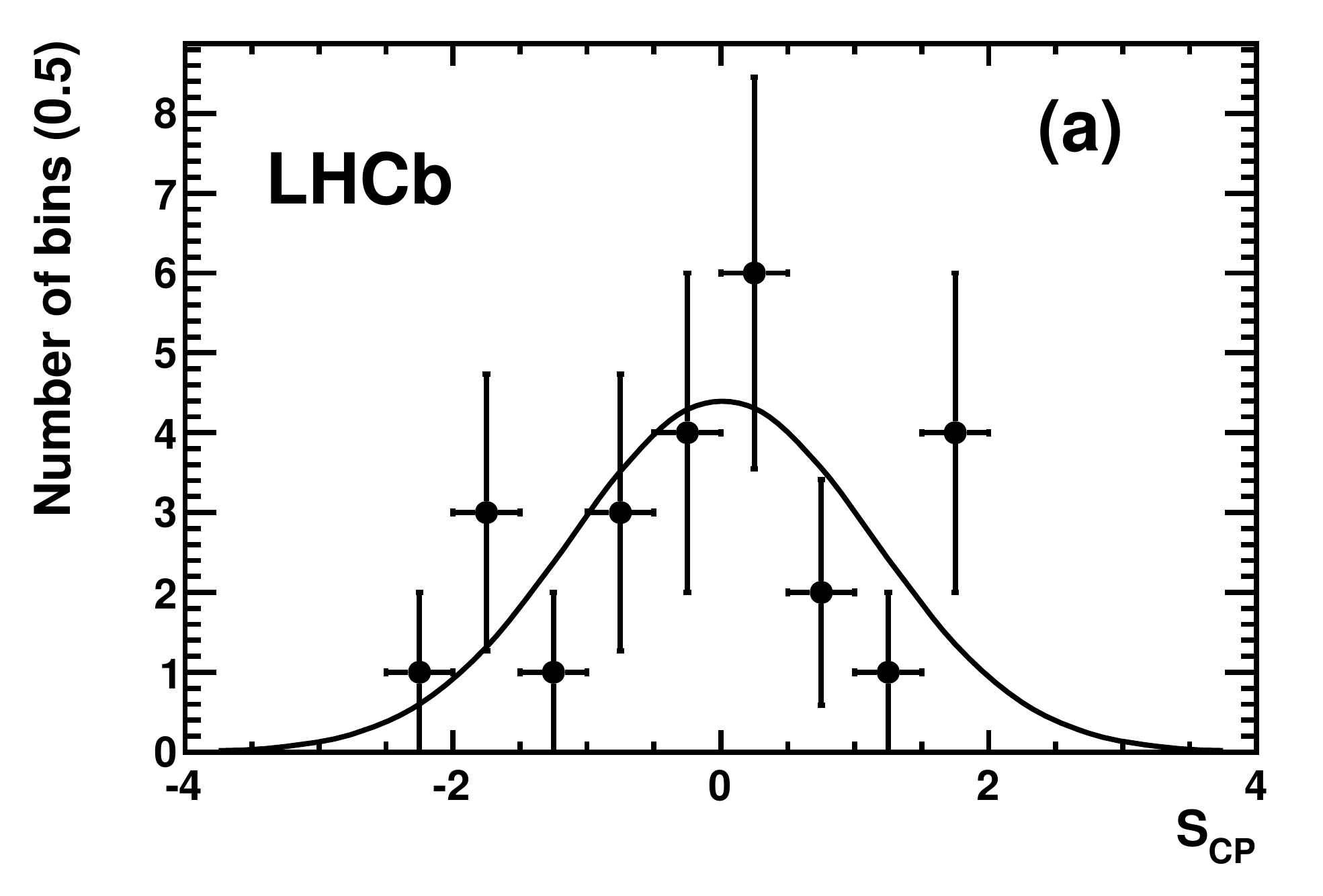}
\includegraphics[width=3.5cm]{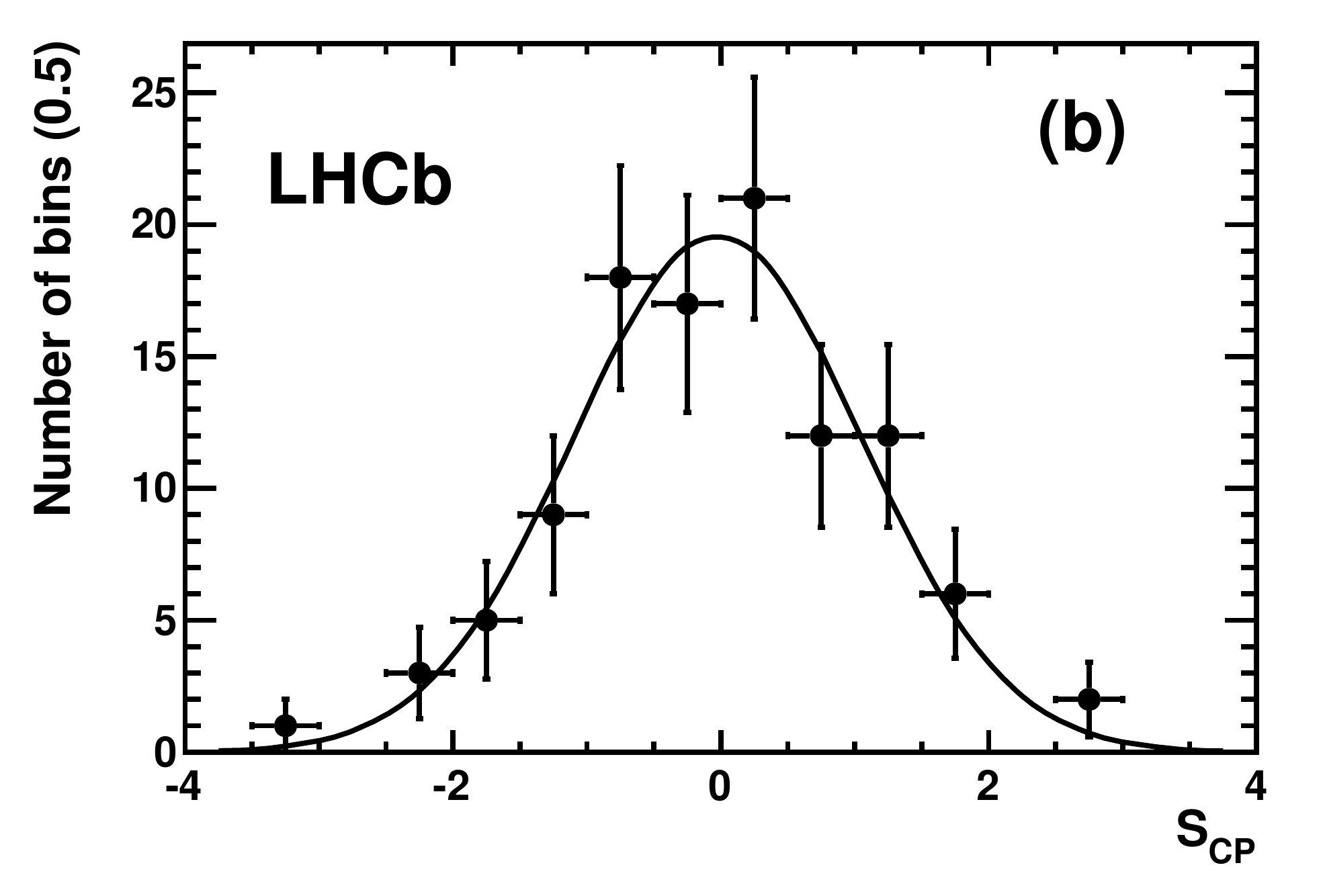}
\includegraphics[width=3.5cm]{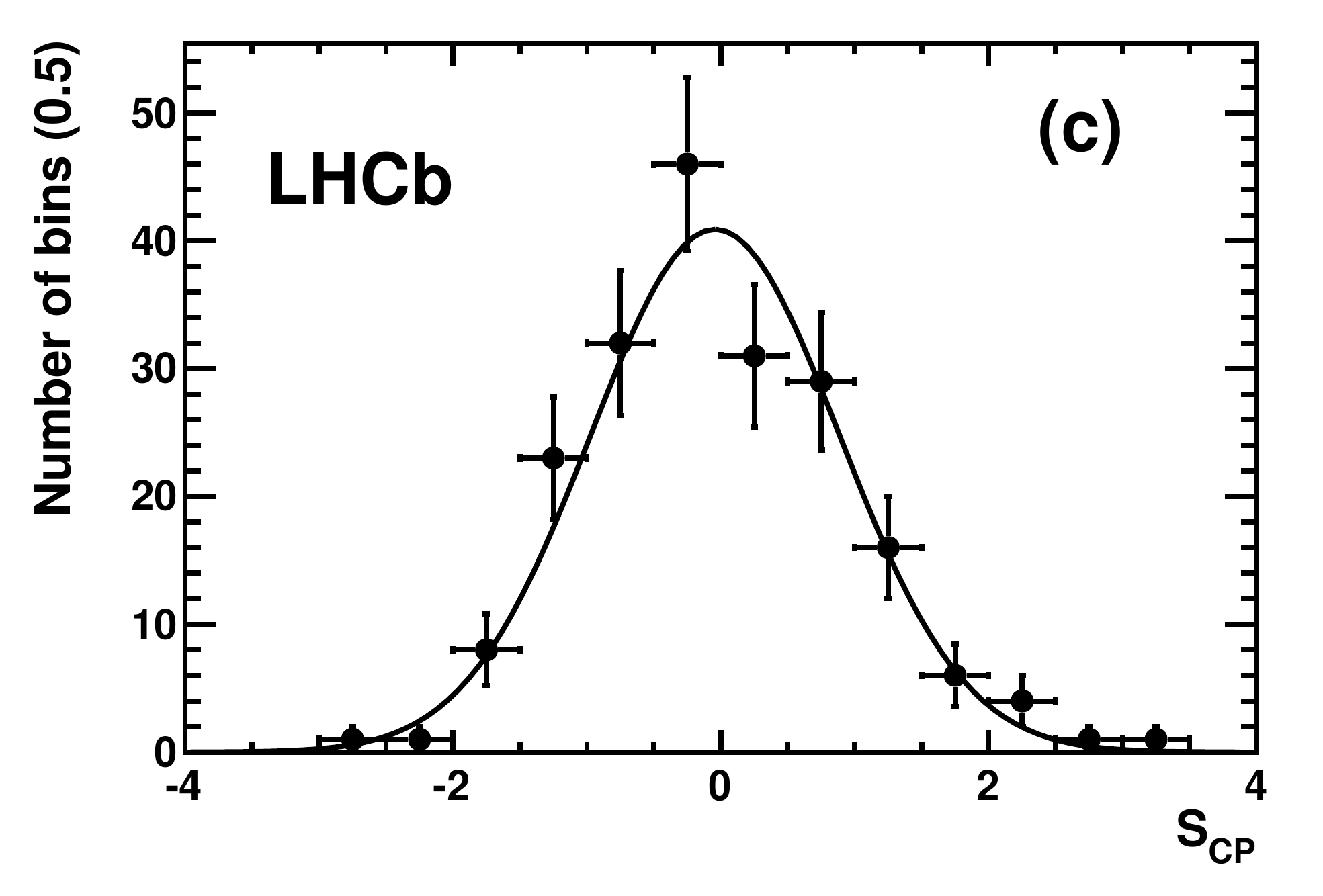}
\includegraphics[width=3.5cm]{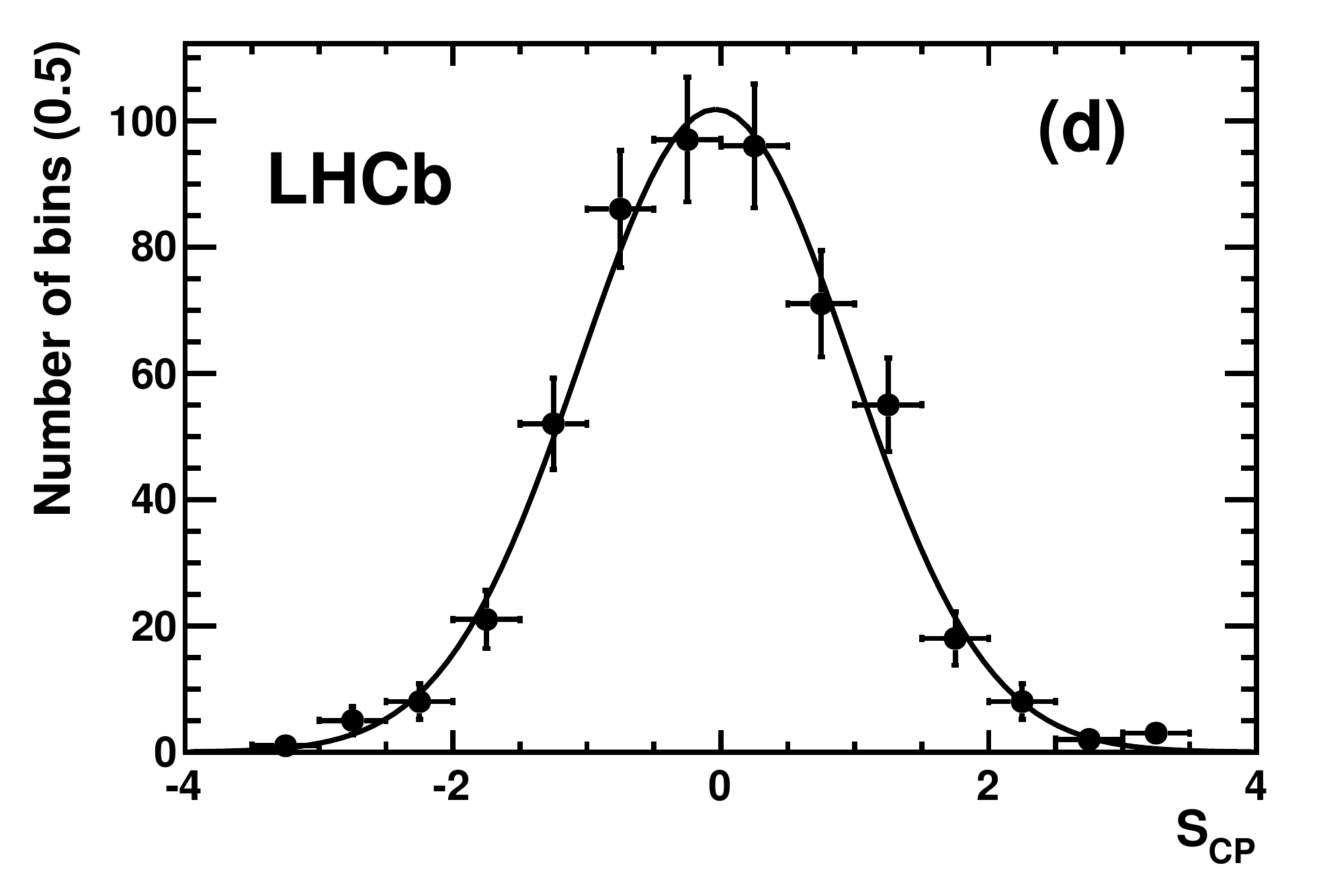}

\caption{Distributions of $S_{CP}$ variable for the LHCb  $D^{+}\to
  K^{-}K^+\pi^+$ Dalitz plot, in phase space (top) and as pulls
  (bottom) for four choices of Dalitz plot binning.}
\label{fig:lb:cp}
\end{figure}

The signal mode is then binned according to four binning schemes, two
of which account for the resonant structure of the decay and two of
which, following the analysis at BaBar \cite{Aubert:2008yd}, do not. The $S_{CP}$ values in
each bin are calculated and the resulting $\chi^{2}$ for consistency
with no CPV is converted into a $p$-value. The $S_{CP}$ values are
shown in Figure~\ref{fig:lb:cp} and the $p$-values are tabulated in
Table~\ref{tab:lb:cp}. No evidence of CPV is seen.

\begin{table}[t]
\begin{center}
\begin{tabular}{l|cc}  
 Binning     &   $\chi^2/{\rm ndf}$  &   $p$-value (\%)  \\ \hline
(a) Adaptive I  &      32.0/24     & 12.7 \\
(b) Adaptive II &    123.4/105    & 10.6    \\ 
(c) Uniform I   & 191.3/198    & 82.1   \\   
(d) Uniform II  &   519.5/529    & 60.5  \\ 
\hline
\end{tabular}
\caption{ Table showing $\chi^{2}/ndf$ and $p$-values for consistency
with no CPV for the binnings shown in Figure~\ref{fig:lb:cp}}
\label{tab:lb:cp}
\end{center}
\end{table}

\section{Four body decays}

The five-dimensional phase space of four body decays presents new
challenges and new opportunities. The binned search for CP violation
can be adapted to four-body decays, as done recently at LHCb in
around 180,000 $D^{0} \to \pi^{+}\pi^{-}\pi^{+}\pi^{+}$ decays \cite{lhcb4pi}. In four-body decays, the method of
$T$-odd moments also becomes available. This has been applied in
several analyses, for example by the BaBar Collaboration in their
search for CPV in $D^{+} \to K^{0}_{S}K^{+}\pi^+\pi^{-}$ ~\cite{Lees:2011dx}.

The proceedure is to define a triple product of the momenta of three
of the particles.
\begin{equation}
C_{T} = \vec{p}_{K^{+}}.(\vec{p}_{\pi^{+}}
  \times \vec{p}_{\pi^{-}})
\end{equation}
The fourth momentum is completely constrained. Then one defines an asymmetry
\begin{equation}
A_{T} = \frac{\Gamma(C_{T} > 0) - \Gamma(C_{T} <
    0)}{\Gamma(C_{T} > 0) + \Gamma(C_{T} < 0)}
\end{equation}
 for $D^{+}$ decays and
  the analogous $A_{\bar{T}}$ for $D^{-}$ decays. The quantity $\mathcal{A} = \frac{1}{2}(A_{T}-A_{\bar{T}})$ is then a measure
  of $T$ violation. $A_{T}$ and $A_{\bar{T}}$ can be individually nonzero if there are
  final state interactions \cite{Bigi:2009zzb}, but their difference cannot, unless CP is
  violated. In a sample of 20,000 decays, BaBar obtain $\mathcal{A} = (-1.2\pm1.0\pm0.46)\%$.

\section{Unbinned techniques}

There is a promising new technique available for unbinned model-independent
searches which has not yet been applied to experimental data~\cite{Williams:2011cd}. It is
possible to compare the densities of oppositely charged points in phase space using a
two-sample test. A test statistic $T$ is used to correlate the
difference between $D^{+}$ and $D^-$ p.d.f.s, $f(\vec{x})$ and
$f(\vec{x'})$, 
\begin{equation}
T = \frac{1}{2}\int\int\left(f(\vec{x}) - \bar{f}(\vec{x})\right)\left(f(\vec{x}') - \bar{f}(\vec{x}')\right)\psi(|\vec{x}-\vec{x}'|)d\vec{x}d\vec{x}'
\end{equation}
To calculate $T$ from data, one then translates this to:
\begin{equation}
T = \frac{1}{n(n-1)}\sum_{i,j>i}^{n} \psi(|\vec{x_{i}}-\vec{x_{j}}|) +
\frac{1}{\bar{n}(\bar{n}-1)}\sum^{\bar{n}}_{i,j>i} \psi(|\vec{x_{i}}-\vec{x_{j}}|)  - \frac{1}{n\bar{n}}\sum_{i, j>i}^{n, \bar{n}}\psi(|\vec{x_{i}}-\vec{x_{j}}|) 
\end{equation}
where $n$ ($\bar{n}$) is the number of events (c.c. events). A
sensible choice for weighting function  $\psi(|\vec{x_{i}}-\vec{x_{j}}|) $ is
  $e^{-(|\vec{x_{i}}-\vec{x_{j}}|)^{2}/2\sigma^{2}}$ where the
  choice of $\sigma$ depends on the scale of the asymmetries expected
  in the Dalitz plot.

The permutation test is then used to determine a $p$-value for
consistency with no CPV. Many datasets where the $``+''$
  and $``-''$ labels are assigned randomly to $D$
  decays are produced, such that there are $n$ $``+''$ decays and $\bar{n}$ $``-''$ decays
  in each set. Then the $p$-value is the fraction of elements in the set for which
  $T > T_{obs}$. 

Preliminary Monte Carlo studies indicate that this method should be
more sensitive than binned searches, although it could also be very
computationally expensive if applied to large datasets in its most
basic form.

\section{Outlook}

Model-independent techniques for CPV searches are likely to
become more important for the very large Belle II and LHCb data samples. Whilst they are not
substitutes for full amplitude analyses, they provide the most
straightforward means to establish CPV in a multi-body decay. It is hoped that model-independent techniques can shed
light on some of the most pressing issues in charm physics in the near future.

\Acknowledgements
I am grateful to the organisers of the Charm 2012 conference and to
the LHCb Collaboration.

\end{document}

%% file: econfmacros.tex
\def\beq{\begin{equation}}
\def\eeq#1{\label{#1}\end{equation}}
\def\eeqn{\end{equation}}

\def\beqa{\begin{eqnarray}}
\def\eeqa#1{\label{#1}\end{eqnarray}}
\def\eeqan{\end{eqnarray}}

\let\bar=\overbar

\def\D{{\cal D}}

\def\Dslash{\not{\hbox{\kern-4pt $D$}}}
\def\dslash{\not{\hbox{\kern-2pt $\del$}}}

\def\msb{{\bar{\ssstyle M \kern -1pt S}}}




%% file: eprint.bbl
\begin{thebibliography}{99}


\bibitem{charmcpv}
  R.~Aaij {\it et al.} [LHCb Collaboration], 
  Phys.\ Rev.\ Lett.\ {\bf 108}, 111602 (2012) 
  [arXiv:1112.0938 [hep-ex]].

\bibitem{Collaboration:2012qw}
  T.~Aaltonen {\it et al.}  [CDF Collaboration],
  Submitted to: Phys. Rev. Lett.
  [arXiv:1207.2158 [hep-ex]].

\bibitem{pdg}
J. Beringer {\it et al.} [Particle Data Group],
Phys.\ Rev.\ D {\bf 86}, 010001 (2012).

\bibitem{Aubert:2008ao} 
  B.~Aubert {\it et al.}  [BABAR Collaboration],
  Phys.\ Rev.\ D {\bf 79}, 032003 (2009)
  [arXiv:0808.0971 [hep-ex]].

 

\bibitem{Aubert:2008yd} 
  B.~Aubert {\it et al.}  [BABAR Collaboration],
  Phys.\ Rev.\ D {\bf 78}, 051102 (2008)
  [arXiv:0802.4035 [hep-ex]].

\bibitem{Bediaga:2009tr} 
  I.~Bediaga, I.~I.~Bigi, A.~Gomes, G.~Guerrer, J.~Miranda and A.~C.~d.~Reis,
  Phys.\ Rev.\ D {\bf 80}, 096006 (2009)
  [arXiv:0905.4233 [hep-ph]].

\bibitem{kkpiLHCb}
  R.~Aaij {\it et al.} [LHCb Collaboration], 
  Phys.\ Rev.\ D.\ {\bf 84}, 112008 (2011) 
  [arXiv:1110.3970 [hep-ex]].

\bibitem{Rubin:2008aa} 
  P.~Rubin {\it et al.}  [CLEO Collaboration],
  Phys.\ Rev.\ D {\bf 78}, 072003 (2008)
  [arXiv:0807.4545 [hep-ex]].

\bibitem{lhcb4pi}
R.~Aaij {\it et al.} [LHCb Collaboration],
LHCb-CONF-2012-019,
Conference note prepared for 36th International Conference on High
Energy Physics, Melbourne, Australia, 4-11 July 2012.

\bibitem{Lees:2011dx} 
  J.~P.~Lees {\it et al.}  [BABAR Collaboration],
  Phys.\ Rev.\ D {\bf 84}, 031103 (2011)
  [arXiv:1105.4410 [hep-ex]].

\bibitem{Bigi:2009zzb}
  I.~Bigi and H.~-B.~Li,
  Int.\ J.\ Mod.\ Phys.\ A {\bf 24S1} (2009) 657.

\bibitem{Williams:2011cd} 
  M.~Williams,
  Phys.\ Rev.\ D {\bf 84}, 054015 (2011)
  [arXiv:1105.5338 [hep-ex]].


\end{thebibliography}
